\documentclass[%
 reprint,
 amsmath,amssymb,
 aps,
]{revtex4-1}

\usepackage{graphicx}
\usepackage{dcolumn}
\usepackage{bm}
\usepackage{dsfont}
\usepackage{appendix}

\usepackage{color}
\usepackage{ulem}

\begin{document}

\preprint{APS/123-QED}

\title{Witten index and dynamical supersymmetry breaking of a gauge theory}

\author{Renata Jora
	$^{\it \bf a}$~\footnote[1]{Email:
		rjora@theory.nipne.ro}}

\affiliation{$^{\bf \it a}$ National Institute of Physics and Nuclear Engineering PO Box MG-6, Bucharest-Magurele, Romania}

\begin{abstract}
We estimate the Witten index for a supersymmetric gauge theory with or without massless matter. For pure supersymmetric Yang Mills our result reproduces  the results in the literature. For the known theories with dynamical symmetry breaking the Witten index is determined to be exactly zero. Our calculation sheds light on the possibility of dynamical supersymmetry breaking  for those theories considered up to now unsolvable.

\end{abstract}
\maketitle

The Witten index \cite{Witten1}, \cite{Witten2}  is defined as:
\begin{eqnarray}
(-1)^F=n_b^0-n_f^0,
\label{wittenidex656}
\end{eqnarray}
where $n_b^0$ is the number of bosonic zero modes and $n_f^0$ is the number of fermionic zero modes. It was shown that for a supersymmetric Yang Mills $SU(N)$ the Witten index is nonzero as it is in general for any supersymmetric gauge theory without matter. For the supersymmetric Yang Mills theory with massive matter one may show that the Witten index is still non zero by extrapolating to the large mass limit where the massive matter can be integrated out.

The Witten index is crucial for studying theories with dynamical supersymmetry breaking because for $(-1)^F=0$ the theory might display this non-perturbative phenomenon.

For $(-1)^F=0$ there are two distinct possibilities:

a) $n_b^0=n_f^0\neq 0$. In this case the supersymmetry is not dynamically broken since there is at least one state with the ground energy zero.

b) $n_f^0=n_b^0=0$.  In this case there is no ground state with
the energy zero hence supersymmetry is dynamically broken.

The Witten index is hard to calculate for many theories, especially for those with massless matter fields that have flat directions. For the latter situation the massless limit cannot be regarded  as the limit $m\rightarrow 0$ of a massive theory as the asymptotic behavior potential may change. Then there might be a discontinuity of the Witten index at the threshold and this  might be even ill-defined.

First it is worth mentioning that for all theories we know that break supersymmetry dynamically the Witten index is zero.
For massless matter theories one needs new methods to analyze the non-perturbative behavior of the theory and various symmetries associated to this. The groundbreaking of Seiberg and his collaborators \cite{Seiberg1}, \cite{Seiberg2} furnished through the power of holomorphicity and dualities an adequate set of tools for elucidating the properties of a supersymmetric gauge theory.  Even so theories that display with certainty dynamical supersymmetry breaking are particular and complicated and may not hold any real connection with what we know up to now about the standard model of elementary particles.

On the other hand there is an alternative treatment of supersymmetric gauge theories \cite{NSVZ} which lead to the famous NSVZ beta function. In \cite{Murayama1} it was shown that by applying Fujikawa methods a direct relation between the holomorphic coupling constant and the canonical one may be obtained. The NSVZ beta function reads:
\begin{eqnarray}
\beta(\alpha)=-\frac{\alpha^2}{2\pi}\frac{[(n_g-\frac{1}{2}n_f)+\frac{1}{2}\sum_{\Psi}\gamma_{\Psi}]}{1-\frac{\alpha}{4\pi}(n_g-n_{\lambda})}.
\label{NSVZ664553}
\end{eqnarray}
Here $n_g$ is the number of gluon zero modes, $n_{\lambda}$ is the number of gluino zero modes, $n_f$ is the number of fermion zero modes, gluino and matter and $\gamma_{\Psi}$ is the anomalous dimension of the matter mass operator. The number of zero modes for $SU(N)$ with $N_f$ multiplets in the fundamental and antifundamental representations was determined to be:
\begin{eqnarray}
&&n_g=4N
\nonumber\\
&&n_{\lambda}=2N
\nonumber\\
&&n_{\Psi}=2N_f,
\label{zeromodes645363}
\end{eqnarray}
where $n_{\Psi}$ is the number of matter fermion zero modes. This leads to a beta function for the $SU(N)$ group:
\begin{eqnarray}
\beta(\alpha)=-\frac{\alpha^2}{2\pi}\frac{3N-N_f(1-\gamma)}{1-\frac{\alpha N}{2\pi}}.
\label{betafunc64536}
\end{eqnarray}

So in one instance we define the zero modes which correspond to the collective coordinates in the instanton approach. On the other hand we define the zero modes in the hamiltonian in a more complex approach. It seems that the two of them are only remotely related. In this work we will show that there might be a connection between the instanton zero modes and those calculated by Witten in \cite{Witten1}. We will  present a method for estimating the Witten index for any theory for which the beta function and the anomalous dimension of the  matter mass operator are known. We will then check our calculations against results in the literature obtained using different methods to find that there is perfect agreement with those.

We consider a supersymmetric gauge theory based on a gauge group $G$ with matter multiplets in arbitrary representations $R_i$ of the gauge group with or without an associated flavor symmetry. The holomorphic Lagrangian of interest is \cite{Murayama1}:
\begin{eqnarray}
&&{\cal L}_h(V_h,\Phi)=\frac{1}{16}\int d^2 \theta\frac{1}{g_h^2}W^a(V_h)W^a(V_h)+h.c+
\nonumber\\
&&\int d^4 \theta \sum_i\Phi_i^{\dagger}\exp[2V_h^i]\Phi_i,
\label{lagrofnterst}
\end{eqnarray}
with the corresponding partition function:
\begin{eqnarray}
Z=\int d V_h d \Phi_i\exp[i\int  d^4x {\cal L}_h ].
\label{partfunc67564}
\end{eqnarray}
The matter fields are considered massless.

There is one more way in which one can write the partition function:
\begin{eqnarray}
Z=\frac{\det\langle F|H|F\rangle}{\det \langle B|H|B\rangle},
\label{secpartfucn64554}
\end{eqnarray}
where $B$ are the bosonic states and $F$ the fermionic ones of the full hamiltonian. This includes zero modes. Thus $Z$ may not be well defined in general but the expression in Eq. (\ref{secpartfucn64554}) makes perfect sense in a finite volume.

In order to go from the Lagrangian in terms of the holomorphic coupling $\frac{1}{g_h^2}$ to that in terms of the canonical coupling $g_c$ one makes the change of variables:
\begin{eqnarray}
&&V_h=g_cV_c
\nonumber\\
&&Z_i^{\frac{1}{2}}\Phi=\Phi',
\label{changeofvariab6465353}
\end{eqnarray}
where $Z_i$ the renormalization constant of the matter wave function.
Then the outcome of this change of variables in the path integral approach is done through the Fujikawa method \cite{Murayama1} and yields new terms in the Lagrangian according to:
\begin{eqnarray}
&&{\cal L}'=\frac{1}{16}\int d^2 \theta \Bigg[[\frac{1}{g_h^2}-\frac{t_G}{8\pi^2}\ln g_c^2+\sum_i\frac{t_{R_i}}{8\pi^2}\ln Z_i]\times
\nonumber\\
&&W^a(g_cV_c)W^a(g_cV_c)+h.c\Bigg]+
\nonumber\\
&&\int d^4 \theta \sum_i\Phi_i^{\prime \dagger}\exp[2V_cg_c]\Phi_i'.
\label{newlagr45665}
\end{eqnarray}
Here $t_G\delta^{ab}={\rm Tr}[T_G^aT_G^b]$ where $T_G$ are the group generators in the adjoint representation and $t_{R_i}\delta^{ab}={\rm Tr}[T_{R_i}^aT_{R_i}^b]$ where $T_{R_i}$ are the group generators in the representation $R_i$.
Then one defines the canonical coupling through the relation:
\begin{eqnarray}
\frac{1}{g_c^2}={\rm Re}\frac{1}{g_h^2}-\frac{t_G}{8\pi^2}\ln g_c^2+\sum_i\frac{t_{R_i}}{8\pi^2}\ln Z_i.
\label{cannn88}
\end{eqnarray}
This leads to the correct correspondence between the beta function for the holomorphic coupling and that for the canonical coupling.

Assume that instead of the change of variables in Eq. (\ref{changeofvariab6465353}) we make the following change of variables:
\begin{eqnarray}
&&V_h=aV_a
\nonumber\\
&&\Phi=a\Phi_a,
\label{resu65774665}
\end{eqnarray}
where $a$ is an arbitrary parameter. Then the partition function as expressed in Eq. (\ref{secpartfucn64554}) will become:
\begin{eqnarray}
&&Z=\frac{\det\langle F_a|H|F_a\rangle}{\det\langle B_a|H|B_a\rangle}=
\nonumber\\
&&\frac{\det\langle F_a|H|F_a\rangle}{\det\langle B_a|H|B_a\rangle}a^{k(n_f^0-n_b^0)},
\label{finalexpr8888}
\end{eqnarray}
where $k$ is a positive finite constant. In the following we will justify the expression in Eq. (\ref{finalexpr8888}). We shall consider the theory in a finite volume where the spectrum of the hamiltonian is discrete and energies are allowed up to some value. Then all the scales in the theories should adequately be smaller that a definite scale. The change of variable in Eq. (\ref{resu65774665}) should not affect the difference between the fermion and boson zero modes. But then in the limit $a\rightarrow 0$ we should obtain the same behavior of the partition function and the same number of zeros. Since the first factor in the second line of Eq. (\ref{finalexpr8888}) brings out $0^{n_f^0-n_b^0}$ then the only possibility is that each zero eigenvalue in the presence of $a$ should be of the type $0\times a^k$ where $k$ is positive, finite and universal.
 The factor $k$ will be determined later by considering an instanton approach.  Finally one can write:
 \begin{eqnarray}
 \frac {d\ln Z}{d \ln a}=k(n_f^0-n_b^0).
 \label{finalexpr775664}
 \end{eqnarray}

On the other hand the change of variables in Eq. (\ref{resu65774665}) can be approached through Fujikawa method using the same method as in \cite{Murayama1}. One can write directly the transformed Lagrangian:
\begin{eqnarray}
&&{\cal L}(aV_a,a\Phi')=
\nonumber\\
&&\frac{1}{16}\int d^2 \theta \Bigg[[\frac{1}{g_h^2}-\frac{t_G}{8\pi^2}\ln a^2-\sum_i\frac{t_{R_i}}{8\pi^2}\ln a^2+
\nonumber\\
&&\sum_i\frac{t_{R_i}}{8\pi^2}\ln Z_i]\times
[W^a(aV_a)W^a(aV_a)+h.c.]\Bigg]+
\nonumber\\
&&\int d^4 \theta a^2\sum_i\Phi_i^{\prime\dagger}\exp[2aV_a]\Phi'.
\label{newlagrty5677}
\end{eqnarray}
Then one determines:
\begin{eqnarray}
&&-i\frac{d\ln Z}{d\ln a}=
\nonumber\\
&&\langle \frac{1}{16}d^2\theta\Bigg[\int d^4 x[\frac{d \frac{1}{g_h^2}}{d\ln a}-\frac{2t_G}{8\pi^2}-\sum_i\frac{2t_{R_i}}{8\pi^2}+
\nonumber\\
&&\sum_i\frac{t_{R_i}}{8\pi^2}\frac{d\ln Z_i}{d\ln a}]\times
[W^a(aV_a)W^a(aV_a)+h.c.]\Bigg]+
\nonumber\\
&&\frac{1}{16}\int d^2 \theta \Bigg[2[\frac{1}{g_h^2}-\frac{t_G}{8\pi^2}\ln a^2-\sum_i\frac{t_{R_i}}{8\pi^2}\ln a^2+
\nonumber\\
&&\sum_i\frac{t_{R_i}}{8\pi^2}\ln Z_i]\times
[W^a(aV_a)W^a(aV_a)+h.c.]\Bigg]+
\nonumber\\
&&\int d^4 \theta 2 a^2\sum_i\Phi_i^{\prime\dagger}\exp[2aV_a]\Phi'\rangle.
\label{compic756647}
\end{eqnarray}
Furthermore one can estimate the contribution of the matter terms as:
\begin{eqnarray}
&&\langle \int d^4 \theta 2 a^2\sum_i\Phi_i^{\prime\dagger}\exp[2aV_a]\Phi'=-i \sum_i\frac{d\ln Z}{d\ln Z_i}=
\nonumber\\
&&\langle \frac{1}{16}d^2\theta\Bigg[\int d^4 xi[\sum_i\frac{2t_{R_i}}{8\pi^2}]\times
\nonumber\\
&&W^a(aV_a)W^a(aV_a)+h.c.\Bigg]\rangle.
\label{mattercontr65765}
\end{eqnarray}
Then  Eqs. (\ref{compic756647}) and (\ref{mattercontr65765}) yield:
\begin{eqnarray}
&&\frac{d\ln Z}{d\ln a}=
\nonumber\\
&&\langle \frac{1}{16}d^2\theta\Bigg[\int d^4 xi[\frac{d \frac{1}{g_h^2}}{d\ln a}-\frac{2t_G}{8\pi^2}+\sum_i\frac{t_{R_i}}{8\pi^2}\frac{d\ln Z_i}{d\ln a}]\times
\nonumber\\
&&W^a(aV_a)W^a(aV_a)+h.c.\Bigg]\rangle+
\nonumber\\
&&\frac{1}{16}\int d^2 \theta \Bigg[2[\frac{1}{g_h^2}-\frac{t_G}{8\pi^2}\ln a^2-\sum_i\frac{t_{R_i}}{8\pi^2}\ln a^2+
\nonumber\\
&&\sum_i\frac{t_{R_i}}{8\pi^2}\ln Z_i]\times
[W^a(aV_a)W^a(aV_a)+h.c.]\Bigg].
\label{finalresult6565}
\end{eqnarray}
Up to a proportionality factor the result in Eq. (\ref{finalresult6565}) represents the Witten index. In order to determine if this cancels or not we need an estimate of the quantity:
\begin{eqnarray}
&&Q=[\frac{d \frac{1}{g_h^2}}{d\ln a}-\frac{2t_G}{8\pi^2}+\sum_i\frac{t_{R_i}}{8\pi^2}\frac{d\ln Z_i}{d\ln a}]+
\nonumber\\
&&2[\frac{1}{g_h^2}-\frac{t_G}{8\pi^2}\ln a^2-\sum_i\frac{t_{R_i}}{8\pi^2}\ln a^2+\sum_i\frac{t_{R_i}}{8\pi^2}\ln Z_i].
\label{finalresultstwo45}
\end{eqnarray}

Finally we can assimilate $a$ with a scale $a=\frac{\mu}{\Lambda}$. Then $Q$ becomes:
\begin{eqnarray}
&&Q=\frac{3t_G-\sum_it_{R_i}}{8\pi^2}-\frac{2t_G}{8\pi^2}+\sum_i\frac{t_{R_i}}{8\pi^2}\gamma_i+
\nonumber\\
&&2[\frac{1}{g_h^2}-\frac{t_G}{8\pi^2}\ln a^2-\sum_i\frac{t_{R_i}}{8\pi^2}\ln a^2+\sum_i\frac{t_{R_i}}{8\pi^2}\ln Z_i]=
\nonumber\\
&&\frac{t_G}{8\pi^2}-\sum_i\frac{t_{R_i}}{8\pi^2}+\sum_i\frac{t_{R_i}}{8\pi^2}\gamma_i+
\nonumber\\
&&2[\frac{1}{g_h^2}-\frac{t_G}{8\pi^2}\ln a^2-\sum_i\frac{t_{R_i}}{8\pi^2}\ln a^2+\sum_i\frac{t_{R_i}}{8\pi^2}\ln Z_i].
\label{expq75665}
\end{eqnarray}
We next consider $a=g_c$.
Using Eq. (\ref{cannn88}) one can further simplify the expression for $Q$ in Eq. (\ref{expq75665}):
\begin{eqnarray}
&&Q=\frac{t_G}{8\pi^2}-\sum_i\frac{t_{R_i}}{8\pi^2}+\sum_i\frac{t_{R_i}}{8\pi^2}\gamma_i+
\nonumber\\
&&2\frac{1}{g_c^2}-\sum_i\frac{t_{R_i}}{8\pi^2}\ln g_c^2.
\label{finalepxrforQ3244}
\end{eqnarray}
We cannot solve Eq. (\ref{finalepxrforQ3244}) in general to see if there is a plausible solution but we can estimate it for some regime when this solution exist. Assume the range of $g_c^2$ where $\frac{g_c^2}{8\pi^2}\geq t_G$. This corresponds to a regime where perturbation theory breaks down and one enters the non-perturbative regime. Then $\frac{1}{g_c^2}$ can be considered small and neglected and a necessary condition for the existence of a solution is:
\begin{eqnarray}
\frac{t_g}{8\pi^2}-\sum_i\frac{t_{R_i}}{8\pi^2}+\sum_i\frac{t_{R_i}}{8\pi^2}\gamma_i\geq0.
\label{cons54663}
\end{eqnarray}
If the coupling constant that is solution to the inequality in Eq. (\ref{cons54663}) is in the domain $\frac{g_c^2}{8\pi^2}\geq t_G$ we say that the condition is also sufficient.

Let us see on a few examples as this might work.

1) $SU(N)$ supersymmetric gauge theory with $N_f$ flavors in the fundamental ($N$) and antifundamental ($N^*$) representations. We need:
\begin{eqnarray}
&&t_R=\frac{1}{2}
\nonumber\\
&&t_G=N
\nonumber\\
&&\gamma=-\frac{N^2-1}{N}\frac{g^2}{8\pi^2}.
\label{firstset34554}
\end{eqnarray}
Then Eq. (\ref{cons54663}) becomes;
\begin{eqnarray}
N-N_f-N_f\frac{N^2-1}{N}\frac{g^2}{8\pi^2}\geq 0,
\label{fisrsconrt65}
\end{eqnarray}
which further leads to,
\begin{eqnarray}
N_f\leq\frac{N}{1+\frac{N^2-1}{N}\frac{g^2}{8\pi^2}}.
\label{first455}
\end{eqnarray}
Condition $\frac{Ng^2}{8\pi^2}\geq 1$ sets the exact limits in Eq. (\ref{first455}) as:
\begin{eqnarray}
N_f\leq\frac{N}{1+\frac{N^2-1}{N^2}}\approx \frac{N}{2}.
\label{firstset3455242}
\end{eqnarray}
Then it is clear that the range $N_f\leq\frac{N}{2}$ for the dynamical supersymmetry breaking coincide with the results in \cite{Jora1} and agree with the general lore regarding this theory.

2) $SO(10)$ with matter in the spinor representation $16$. The group factors are:
\begin{eqnarray}
&&t_R=2
\nonumber\\
&&t_G=N-2=8
\nonumber\\
&&\gamma=\frac{45}{4}\frac{g^2}{8\pi^2}.
\label{secondcase23}
\end{eqnarray}

Then condition (\ref{cons54663}) becomes:
\begin{eqnarray}
6-\frac{45}{2}\frac{g^2}{8\pi^2}\geq 0,
\label{second45553}
\end{eqnarray}
which translates into:
\begin{eqnarray}
\frac{g^2}{8\pi^2}\leq\frac{12}{45},
\label{second6299}
\end{eqnarray}
which is in the range $\frac{g^2}{8\pi^2}\geq\frac{1}{8}$ which means that supersymmetry is dynamically broken. This results was obtained through a different approach in \cite{Murayama2}.

3) $SU(5)$ with matter in the antisymmetric  ($10$) and antifundamental $5*$ representations:

We need only the structure  factor for the antisymmetric representation $t_{anti}=\frac{3}{2}$. Condition (\ref{cons54663}) becomes:
\begin{eqnarray}
3-10\frac{g^2}{8\pi^2}\geq 0,
\label{third54663}
\end{eqnarray}
which further leads to:
\begin{eqnarray}
\frac{g^2}{8\pi^2}\leq\frac{3}{10},
\label{res65775}
\end{eqnarray}
which is in the range $\frac{g^2}{8\pi^2}\geq\frac{1}{5}$. This means that supersymmetry is dynamically broken which again agrees with the results in \cite{Murayama2}.

4) $SU(N)$ with matter in the symmetric ($\frac{N(N+1)}{2}$) and antifundamental representations. For the symmetric representation $t_{sym}=\frac{N+2}{2}$ and $C_F=N-\frac{2}{N}+1$.

 Condition (\ref{cons54663}) becomes:
\begin{eqnarray}
\frac{N}{2}-\frac{3}{2}-\frac{g^2}{8\pi^2}[\frac{N}{2}+(N+1-\frac{2}{N})(N+2)]\geq 0.
\label{finalex}
\end{eqnarray}
This further leads to:
\begin{eqnarray}
\frac{g^2}{8\pi^2}\leq\frac{N-3}{N+2(N+1-\frac{2}{N})(N+2)}\leq\frac{1}{N}.
\label{finalex6455}
\end{eqnarray}
This shows that the Witten index cannot be zero and that supersymmetry cannot be dynamically broken. Again the results agrees with the findings in the literature \cite{Seiberg3}.

In this work we estimated up to factor of proportionality the Witten index for a supersymmetric gauge theory. If one admits \cite{Jora2} that whenever the Witten index is zero the supersymmetry is dynamically broken then we proposed a clear criterion for dynamical supersymmetry breaking.  This criterion depends on the beta function and on the anomalous dimension of the matter mass operators but is independent on the specific non-perturbative dynamics.  We verified the validity of our result for four supersymmetric gauge theories for which whether the supersymmetry is dynamically broken or not is known through other methods. Our results agree in totality with those in the literature.

In summary we presented a method for calculating the Witten index  and hence introduced a necessary and sufficient condition for dynamical supersymmetry breaking that can be applied  to all supersymmetric gauge theories, with or without flat directions, calculable or non calculable. Our findings may have far reaching and important applications.


\begin{thebibliography}{30}


\bibitem{Witten1} E. Witten, Nucl. Phys. B {\bf 202}, 253-316 (1982).
\bibitem{Witten2} E. Witten, Nucl. Phys. B {\bf 185}, 513-554 (1981).
\bibitem{NSVZ} V. A. Novikov, M. A. Shifman, A. I. Vainshtein and V. I. Zakharov Nucl. Phys. B {\bf 229}, 381 (1983);
Phys. Lett. B {\bf 139}, 389 (1984); Nucl. Phys. B {\bf 260}, 157 (1985); Phys. Lett. B {\bf 166}, 329 (1986).
\bibitem{Murayama1} N. Arkani-Hamed and H. Murayama, JHEP {\bf 0006}, 030 (2000).
\bibitem{Seiberg1} N. Seiberg, Nucl. Phys. B {\bf 435}, 129 (1995).
\bibitem{Seiberg2} K. A. Intriligator and N. Seiberg, Nucl. Phys. Proc. Suppl. {\bf 45} BC:1-28 (1996).
\bibitem{Murayama2} H. Murayama, Phys. Lett. B {\bf 355}, 187-192 (1995).
\bibitem{Jora1} R. Jora, arXiv:1902.11076 (2019).
\bibitem{Jora2} R. Jora, arXiv:1901.01539 (2019).
\bibitem{Seiberg3} I. Affleck, M. Dine and N. Seiberg, Nucl. Phys. B {\bf 256}, 557-599 (1985).










\end{thebibliography}
\end{document}